\documentstyle[sprocl]{article}

\input{psfig}

\def\Journal#1#2#3#4{{#1} {\bf #2}, #3 (#4)}

\def\PRD{{\em Phys. Rev.} D}

\def\be{\begin{equation}}
\def\ee{\end{equation}}
\def\bea{\begin{eqnarray}}
\def\eea{\end{eqnarray}}
\begin{document}

\title{ELECTRICAL CONDUCTION IN THE EARLY UNIVERSE}

\author{H. HEISELBERG}

\address{Nordita, Blegdamsvej 17, 2100 Copenhagen \O, Denmark\\E-mail: hh@nordita.dk} 

\maketitle\abstracts{The electrical conductivity has been 
calculated \cite{BH} in the early universe at
temperatures below as well as above the electroweak vacuum scale, $T_c\simeq
100$GeV.  Debye and dynamical screening of electric and magnetic interactions
leads to a finite conductivity, $\sigma_{el}\sim T/\alpha\ln(1/\alpha)$, at
temperatures well below $T_c$.  At temperatures above, $W^\pm$ charge-exchange
processes -- analogous to color exchange through gluons in QCD -- effectively
stop left-handed charged leptons.  However, right-handed leptons can carry
current, resulting in $\sigma_{el}/T$ being only a factor $\sim
\cos^4\theta_W$ smaller than at temperatures below $T_c$.}

\section{Introduction}
    The strong magnetic fields measured in many spiral galaxies, $B\sim
2\times 10^{-6}$ G, are conjectured to be produced primordially;
proposed mechanisms include fluctuations during an inflationary universe
\cite{Turner} or at the GUT scale, and plasma turbulence during the
electroweak transition or in the quark-gluon
hadronization transition.  The production and later
diffusion of magnetic fields depends crucially on the electrical conductivity,
$\sigma_{el}$, of the matter in the universe; typically, over the age of the
universe, $t$, fields on length scales smaller than $L\sim
(t/4\pi\sigma_{el})^{1/2}$ are damped. 

    The electrical conductivity was estimated in \cite{Turner} in the
relaxation time approximation as $\sigma_{el}\sim n\alpha\tau_{el}/m$ with
$m\sim T$ and relaxation time $\tau_{el}\sim 1/(\alpha^2T)$, where $\alpha =
e^2/4\pi$.  In Refs.  \cite {HK,Enqvist} the relaxation time was
corrected with the Coulomb logarithm.  A deeper understanding of the screening
properties in QED and QCD plasmas has made it possible to
calculate a number of transport coefficients including viscosities, diffusion
coefficients, momentum stopping times, etc., exactly in the weak coupling
limit \cite{BP,tran,eta,GT}.  However, calculation of processes
that are sensitive to very singular forward scatterings remain problematic.
For example, the calculated color diffusion and conductivity \cite{color},
even with dynamical screening included, remain infrared divergent due to color
exchange in forward scatterings.  Also the quark and gluon damping rates at
non-zero momenta calculated by resumming ring diagrams exhibit infrared
divergences \cite{damp} whose resolution requires more careful analysis
including higher order diagrams as, e.g., in the Bloch-Nordsieck calculation
of Ref.  \cite{Blaizot}.  Charge exchanges through $W^\pm$, processes
similar to gluon color exchange in QCD, are important in forward scatterings
at temperatures above the $W$ mass, $M_W$. 

\section{Electrical conductivities in high temperature QED}

    The electrical conductivity in the electroweak symmetry-broken
phase at temperatures below the electroweak boson mass scale,
$M_W\gg T\gg m_e$, is dominated by charged leptons
$\ell=e^-,\mu^-,\tau^-$ and anti-leptons $\bar{\ell}=e^+,\mu^+,\tau^+$
currents.  In the broken-symmetry phase weak interactions between
charged particles, which are generally smaller by a factor
$\sim(T/M_W)^4$ compared with photon-exchange processes, can be
ignored.  The primary effect of strong interactions is to limit the
drift of strongly interacting particles, and we need consider only
electromagnetic interactions between charged leptons and quarks.

    Transport processes are most simply described by the Boltzmann kinetic
equation for the distribution functions of particle
species $i$, of charge $e$. We use the standard methods of linearizing
around equilibrium with a collision term with interactions given by
perturbative QED. We refer to \cite{BH} for details of the calculation
but note the essential physics of Debye screening of the longitudinal
(electric) interactions and Landau damping of the transverse (magnetic)
interactions.
  The electrical conductivity for charged leptons is 
to leading logarithmic order
\cite{tran,BH}
\begin{eqnarray}
   \sigma_{el}^{(\ell\bar{\ell})} &\equiv& j_{\ell\bar{\ell}}/E =\,
\frac{3\zeta(3)}{\ln2} \frac{T}{\alpha\ln(C/\alpha N_l)}\,,
     \quad m_e\ll T\ll T_{QGP}.
     \label{ec}
\end{eqnarray}
where the constant $C\sim 1$ in the logarithm gives the next to leading order
terms\cite{Ahonen,eta}.
Note that the number of lepton species drops out except in the logarithm.
The above calculation taking only electrons as massless leptons ($N_l=1$)
gives a first approximation to the electrical conductivity in the temperature
range $m_e\ll T\;\ll\; T_{QGP}$, below the hadronization transition,
$T_{QGP}\sim 150$ GeV, at which hadronic matter undergoes a transition to a
quark-gluon plasma.  Thermal pions and muons in fact also reduce the
conductivity by scattering electrons, but they do not become significant
current carriers because their masses are close to $T_{QGP}$.

    For temperatures $T > T_{QGP}$, the matter consists of leptons and
deconfined quarks.  The quarks themselves contribute very little to the
current, since strong interactions limit their drift velocity.
However, they
are effective scatterers, and thus modify the conductivity
(see \cite{tran,BH} and Fig. 1):
\begin{eqnarray}
\sigma_{el} =
    \frac{N_l}{N_l+ 3 \sum_q^{N_q} Q^2_q}\sigma_{el}^{(\ell\bar{\ell})}
  , \quad       T_{QGP}\ll\; T\;\ll\; M_W.
\label{eq}
\end{eqnarray}

\begin{figure}[t]
\psfig{figure=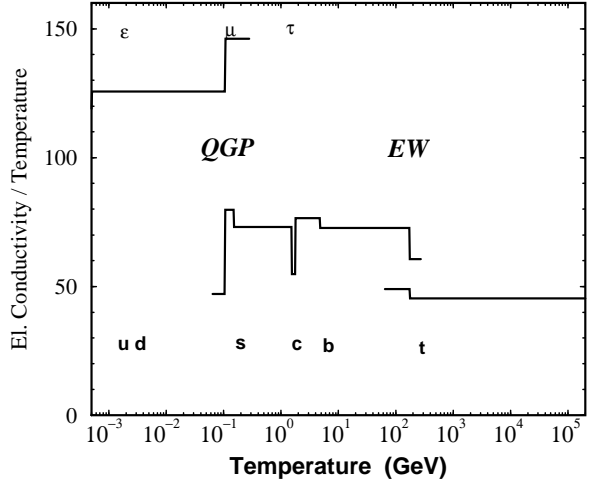,height=2.9in,angle=-90}
\caption{The ratio of the electrical conductivity to temperature,
$\sigma_{el}/T$ vs. temperature.  The
temperatures, where the transitions from hadronic to quark-gluon plasma and
electroweak symmetry breaking occur, are indicated by QGP and EW respectively.
The conductivity $\sigma_{el}$ is given by Eqs.  (\ref{ec},\ref{eq},\ref{sTT})
in the three regions and are extrapolated into the regions of the phase
transitions.  The quark and lepton masses in the figure indicate the
temperatures at which they are thermally produced and thus affect the
conductivity. 
For clarity, the quarks and leptons are assumed to make their appearance
abruptly when $T$ becomes $>m_i$; in reality they are gradually
produced thermally as $T$ approaches their masses.
 \label{fig:el}}
\end{figure}

\section{The symmetry-restored phase}

    To calculate the conductivity well above the electroweak transition, $T\gg
T_c\simeq 100$GeV \cite{Kajantie}, 
where the electroweak symmetries are fully restored, we describe the
electroweak interactions by the standard model Weinberg-Salam Lagrangian with
minimal Higgs ($\phi$).
  At temperatures below $T_c$, the Higgs mechanism
naturally selects the representation $W^\pm$, $Z^0$, and $\gamma$ of the four
intermediate vector bosons.  At temperatures above the transition -- where
$\langle\phi\rangle$ vanishes for a sharp transition, or tends to zero for a
crossover -- we consider driving the system with external vector potentials
$A^a=B,W^\pm,W^3$, which give rise to corresponding ``electric'' fields
$E_i^a \equiv F_{i0}^a = \partial_i A_0^a - \partial_0 A_i^a$ for $A^a=B$ and
$F_{i0}^a = \partial_i A_0^a - \partial_0 A_i^a
 -g \epsilon_{abc}A_i^bA_0^c$ for $A^a=W^1,W^2,W^3$.
One can equivalently drive the system with the electromagnetic and weak
fields derived from $A$, $Z^0$, and $W^\pm$, as when $T\ll T_c$, or any other
rotated combination of these.  We consider here only the weak field limit and
ignore the nonlinear driving terms.  The self-couplings
between gauge bosons are important, however, in the scattering processes in
the plasma determining the conductivity.

    The electroweak fields $A^b$ act on the matter to generate currents $J_a$
of the various particles present in the plasma, such as left and
right-handed leptons and their antiparticles, and quarks, vector bosons, and
Higgs bosons.  The Higgs and vector boson contributions are
negligible. Therefore the significant terms in the currents are
$J_B^\mu = \frac{g'}{2} (\bar{L}\gamma^\mu Y L+\bar{R}\gamma^\mu Y R)$
and $J_{W^i}^\mu = \frac{g}{2} \bar{L}\gamma^\mu \tau_iL$.
We define the conductivity tensor $\sigma_{ab}$ in general by
\begin{eqnarray}
    {\bf J}_a = \sigma_{ab} {\bf E}^b. \label{sdef}
\end{eqnarray}

    The electroweak $U(1)\times SU(2)$ symmetry implies that the conductivity
tensor, $\sigma_{ab}$, in the high temperature phase is diagonal in the
representation , as can be seen directly from the (weak field) Kubo formula
which relates the conductivity to (one-boson irreducible) current-current
correlation functions.
The construction of the conductivity in terms of the Kubo formula
assures that the conductivity and hence the related entropy production
in electrical conduction are positive.  Then
$\sigma =Diag(\sigma_{BB},\sigma_{WW},\sigma_{WW},\sigma_{WW})$.
Due to isospin symmetry of the $W$-interactions the conductivities
$\sigma_{W^iW^i}$ are the same, $\equiv\sigma_{WW}$, but differ from the
$B$-field conductivity, $\sigma_{BB}$.

    The calculation of the conductivities $\sigma_{BB}$ and $\sigma_{WW}$ in
the weak field limit parallels that done for $T\ll T_c$.  The main difference
is that weak interactions are no longer suppressed by a factor $(T/M_W)^4$ and
the exchange of electroweak vector bosons must be included.  The conductivity,
$\sigma_{BB}$, for the abelian gauge field $B$ can be calculated similarly to
the electrical conductivity at $T\ll T_c$. 

    Although the quarks and $W^\pm$ are charged, their drifts in the presence
of an electric field do not significantly contribute to the electrical
conductivity.  Charge flow of the quarks is stopped by strong interactions,
while similarly flows of the $W^\pm$ are effectively stopped by $W^+ + W^- \to
Z^0$, via the triple boson coupling.  Charged Higgs bosons are likewise
stopped via $W^\pm\phi^\dagger\phi$ couplings.  These particles do, however,
affect the conductivity by scattering leptons.
Particle masses have negligible effect on the conductivities\cite{BH}.

    These considerations imply that the $B$ current consists primarily of
right-handed $e^\pm$, $\mu^\pm$ and $\tau^\pm$, interacting only through
exchange of uncharged vector bosons $B$, or equivalently $\gamma$ and $Z^0$.
Because the left-handed leptons interact through ${\bf W}$ as well as through
$B$, they give only a minor contribution to the current.  They are, however,
effective scatterers of right-handed leptons.  The resulting conductivity is
(see \cite{BH} for details)
\begin{eqnarray}
   \sigma_{BB} &=& \frac{9}{19} \cos^2\theta_W\sigma_{el}
     \,,\quad T\gg T_c \,.     \label{sTT}
\end{eqnarray}

    Applying a $W^3$ field to the electroweak plasma drives the charged
leptons and neutrinos oppositely since they couple through $g\tau_3W_3$.  In
this case, exchanges of $W^\pm$ dominate the interactions as charge is
transferred in the singular forward scatterings and 
Landau damping is not sufficient to screen the interaction,
a QCD magnetic (gluon) mass, $m_{mag}\sim g^2T$, 
will provide an infrared cutoff. We expect that
$\sigma_{WW} \sim \alpha\, \sigma_{BB}$.
This effect of $W^\pm$ exchange is analogous to the way gluon exchange in
QCD gives strong stopping and reduces the ``color conductivity" significantly
\cite{color}; similar effects are seen in spin diffusion in Fermi liquids
\cite{BP}.

    The electrical conductivity is found from $\sigma_{BB}$ and $\sigma_{WW}$
by rotating the $B$ and $W^3$ fields and currents by the Weinberg angle; using
Eq.  (\ref{sdef}),
$(J_A,J_{Z^0})={\cal R}(\theta_W)\sigma {\cal R}(-\theta_W) (A,Z^0)$.
Thus the electrical conductivity is given by
\begin{eqnarray}
   \sigma_{AA} =
        \sigma_{BB}\cos^2\theta_W + \sigma_{WW}\sin^2\theta_W;
\end{eqnarray}
$\sigma_{el}/T$ above the electroweak transition differs from that below
mainly by a factor $\sim\cos^4\theta_W\simeq 0.6$.

\section{Summary and Outlook}

Typically,
$\sigma_{el}\simeq T/\alpha\ln(1/\alpha)$, where the logarithmic dependence
on the coupling constant arises from Debye and dynamical screening of small
momentum-transfer interactions.  In the quark-gluon plasma, at $T\gg
T_{QGP}\sim 150$ MeV, the additional stopping on quarks reduces the electrical
conductivity from that in the hadronic phase.  In the electroweak
symmetry-restored phase, $T\gg T_c$, interactions between leptons and $W^\pm$
and $Z^0$ bosons reduce the conductivity further.  It
does not vanish (as one might have imagined to result from singular unscreened
$W^\pm$-exchanges), and is larger than previous estimates, within an order of
magnitude.  The current is carried mainly by right-handed leptons since they
interact only through exchange of $\gamma$ and $Z^0$.

    From the above analysis we can infer the qualitative behavior of
other transport coefficients.  The characteristic electrical
relaxation time, $\tau_{el}\sim (\alpha^2\ln(1/\alpha)T)^{-1}$,
defined from $\sigma_{el}\simeq e^2 n\tau_{el}/T$, is a typical
``transport time" which determines relaxation of transport processes
when charges are involved.  Right-handed leptons interact through
$Z^0$ exchanges only, whereas left-handed leptons may change into
neutrinos by $W^\pm$ exchanges as well.  Since $Z^0$ exchange is
similar to photon exchange when $T\gg T_c$, the characteristic
relaxation time is similar to that for electrical conduction,
$\tau_{\nu}\sim (\alpha^2\ln(1/\alpha)T)^{-1}$ (except for the
dependence on the Weinberg angle).  Thus the viscosity is $\eta\sim
\tau_\nu \sim T^3/(\alpha^2\ln(1/\alpha))$.  For $T\ll M_W$ the
neutrino interaction is suppressed by a factor $(T/M_W)^4$; in this
regime neutrinos have longest mean free paths and dominate the
viscosity.

    The electrical conductivity of the plasma in the early universe is
sufficiently large that large-scale magnetic flux present in this period does
not diffuse significantly over timescales of the expansion of the universe.
The time for magnetic flux to diffuse on a distance scale $L$ is $\tau_{diff}
\sim \sigma_{el} L^2$.  Since the expansion timescale $t_{exp}$ is $\sim
1/(t_{\rm Planck}T^2)$, where $t_{\rm Planck} \sim 10^{-43}$ s is the Planck
time, one readily finds that
$\frac{\tau_{diff}}{t_{exp}} \sim
\alpha x^2 \frac{\tau_{el}}{t_{\rm Planck}} \gg 1$,
where $x = L/ct_{exp}$ is the diffusion length scale in units of the
distance to the horizon.  As described in Refs.  \cite{Enqvist,LMcL}, 
sufficiently large domains with magnetic fields in the early
universe would survive to produce the primordial magnetic fields observed
today.

\section*{Acknowledgments}
To my collaborator G. Baym, and for discussions with C.J. Pethick and J.Popp.

\end{document}